\documentclass{article}
\usepackage[T1]{fontenc}
\usepackage{graphicx}
\usepackage{epsf,rotate}
\usepackage{epsfig}
\usepackage{rotating}
\makeatletter

\newcommand{\LyX}{L\kern-.1667em\lower.25em\hbox{Y}\kern-.125emX\spacefactor1000}

\usepackage[T1]{fontenc}

\makeatletter

\usepackage[T1]{fontenc}
\usepackage{geometry}


\makeatother

\begin{document}

\title{A Model for Damage Spreading with Damage Healing:
Monte Carlo Study of the two Dimensional Ising Ferromagnet.}
\author{M. Leticia Rubio Puzzo and Ezequiel V. Albano}
\date{Instituto de Investigaciones Fisicoqu\'{\i}micas Te\'{o}ricas 
y Aplicadas (INIFTA), UNLP, CONICET, 
Casilla de Correo 16 Sucursal 4, (1900) La Plata, Argentina.}
%\date{today}
\maketitle

\begin{abstract}
An Ising  model for damage spreading with a probability of damage 
healing ($q = 1 - p$) is proposed and studied by means of Monte 
Carlo simulations. In the limit $p \rightarrow 1$ the new model is 
mapped to the standard Ising model. It is found that, for temperatures 
above the Onsager critical temperature ($T_{C}$), there exist a 
no trivial finite value of $p$ that sets the critical point 
($p_{c}$) for the onset of damage spreading. It is found that $p_{c}$ 
depends on $T$, defining a critical curve at the border between 
damage spreading and damage healing. Transitions along such curve are 
found to belong to the universality class of directed percolation. 
The phase diagram of the model is also evaluated showing that for 
large $T$ one has $p_{c} \propto (T- T_{C})^{\alpha}$, with $\alpha = 1$. 
Within the phase where the damage remains active, the stationary value 
of the damage depends lineally on both $p - p_{c}$ and $T - T_{C} $.
\end{abstract}

\vskip 1.0 true cm
	

\vskip 1.0 true cm


\pagebreak

\section*{I. Introduction}

The damage spreading method has been widely used to study the critical
properties of Ising-like systems 
\cite{derrS,stan,grass1,montani,wang,vojta,lima1,lima2} as well 
as to spin glass \cite{derrW,argo}. The method is based on the
synchronous Monte Carlo update of two distinct spin configurations
that are evolving from an almost identical initial state \cite{herr}. 
Since the Ising model lacks of an intrinsic dynamic, it has to be 
chosen a particular one. Among others, the more frequently used 
are Glauber and heat-bath dynamics. In contrast to usual statistical
Monte Carlo studies, damage spreading results depends on the used
dynamic \cite{herr}. This behavior has been used to evaluate the
dynamical exponent $z$ of the 2D and 3D Ising models with
different dynamics \cite{grass1,wang1,sta}. Also, very recently, 
it has been shown that damage spreading is a powerful and useful
technique for the numerical study of the role of the interfaces 
between magnetic domains on the propagation of 
perturbations in magnetic materials \cite{we1,we2}. 
Within this context, we have reported that the presence of interfaces 
act as a ''catalyst'' of the damage in at least two different ways: 
speeding up the propagation and causing an enhancement of the 
total damaged area \cite{we1,we2}.

The basic idea for the implementation of the damage 
spreading method \cite{herr} 
is to start from an equilibrium configuration
of the system at temperature $T$, which is generically called \( S^{A}(T) \).
Subsequently, a very small perturbation is applied to that configuration 
in order to obtain a new one, i.e. the so called perturbed 
configuration \( S^{B}(T) \). Usually the perturbed configuration 
is obtained by flipping a small number of spins of the unperturbed 
configuration. Then one has to study the time evolution of the 
perturbation in order to investigate under which conditions such 
a small perturbation will grow up indefinitely or
eventually (hopefully) it will vanish and become healed.
In order to follow the time evolution of the perturbation an useful method is
to measure the ``Hamming distance'' or damage between the unperturbed
and the perturbed configurations. The total damage \( D(t) \) is defined as
the fraction of spins with different orientations, that is

\begin{equation}
D(t)=\frac{1}{2N}\sum ^{N}_{l}\left| S^{A}_{l}(t,T)-S^{B}_{l}(t,T)]\right|,
\label{eq:dam}
\end{equation}

\noindent where the summation runs over the total number of spins N and the 
index \( l\, (1\leq l\leq N) \) is the label that identifies the 
spins of the configurations.  Starting from a vanishing small 
perturbation \( D(t=0)\rightarrow 0 \) one can expects at least 
two scenarios, namely: a) \( D(t\rightarrow \infty )\rightarrow 0 \)
and the perturbation is irrelevant because it become healed 
and b) \( D(t\rightarrow \infty ) \) goes to some well defined finite 
value. In the latter, frequently undesired case, the perturbation is 
relevant because it can not be healed out. For further details see the 
review \cite{herr}.

The aim of this work is propose and study a model for damage spreading 
with damage healing, which is based in the two dimensional ($2D$) Ising model.
It is well known that the $2D$ Ising magnet undergoes a second-order
order-disorder transition when the temperature is raised from a relatively low
initial value. The location of the critical point is know exactly and it is
given by the so called Onsager critical 
temperature \( \frac{kT_{C}}{J}=2.269... \).
According to previous studies of damage spreading using Glauber dynamics 
it is known that there is a critical damage temperature given by 
\( T_{D}\cong 0.992\, T_{C} \) \cite{grass2,grass3}, such as 
for \( T>T_{D} \) the damage spreads out over the whole sample while 
for \( T<T_{D} \) the damage becomes healed after some finite time.
Consequently, the proposed model, that incorporates a healing probability,
is suitable for the study of damage spreading above \( T_{D} \), as well 
as to gain insight on the robustness of the damage behavior in the 
Ising model.

The manuscript is organized as follows: in Section II we propose the 
model and describe the numerical procedure for the simulation of 
damage spreading. Section III is devoted to the presentation and 
discussion of the results, while our conclusions are stated in Section IV.

\section*{II. A Model for Damage Spreading with Healing and the 
Monte Carlo Simulation Method.}

The model is based on the $2D$ ferromagnetic Ising model, which for a
square lattice of side $L$ can 
be described by the following Hamiltonian \( H \):

\begin{equation}
H = -J \sum ^{L,L}_{<ij,mn>} \sigma _{ij}\sigma _{mn}
\label{eq:ham}
\end{equation}

\noindent where \( \sigma _{ij} \) is the spin variable, 
corresponding to the site
of coordinates (i, j), that may assume two different values, 
namely \( \sigma _{ij}=\pm 1 \), \( J>0 \) is the coupling constant of 
the ferromagnet and the summation of (\ref{eq:ham}) runs over all 
the nearest-neighbor pairs of spins.

For the purpose of the simulations, periodic boundary conditions are 
always used. Furthermore, we have used  the Glauber dynamics. In 
order to implement this dynamic a randomly selected spin is flipped 
with probability \( p(flip) \) given by:

\begin{equation}
p(flip)=\frac{\exp (-\beta \cdot \bigtriangleup H)}{1+\exp (-\beta \cdot \bigtriangleup H)}
\label{eq:rates}
\end{equation}

\noindent where \( \bigtriangleup H \) is the difference between the 
energy of the would-be new configuration and the old configuration, 
and \( \beta =1/kT \) is the usual Boltzmann factor. The temperature 
is measured in units of the Onsager critical temperature of the $2D$
Ising model. 

In order to study the time evolution of the damage spreading a meaningful 
definition of the Monte Carlo time step (mcs) is necessary. For this 
purpose the standard definition is adopted according to that during one 
mcs all \( L\times L \) spins of the sample are flipped once, in the average.

For the practical implementation of equation (\ref{eq:dam}), first 
an equilibrium configuration (say configuration \( S^{A}_{e} \)) 
is generated. For this purpose one starts from an initial configuration 
with all \( N = L\times L\) spins oriented at random. The application 
of the Glauber dynamics leads to the desired equilibrium
configuration after \( 10^{4} \)mcs. Subsequently, a fully damaged 
replica \( S^{B}_{e} \) of such configuration is created, such as 
\( S^{B}_{e} \) is the mirror image of  \( S^{A}_{e} \) and consequently
$D(0) = 1$. Of course, due to the spin flip symmetry, the replica
configuration is also equilibrated.

In standard damage spreading studies, the standard Monte Carlo procedure
is then applied to both configurations, where the same sites are randomly 
selected and the same random numbers are used in both systems in order 
to perform the updates. For the purpose of the proposed model we first 
define the healing probability $q$ and we used the
same method but according the following rules: 

i) A sample site of coordinates $(i,j)$ is selected at random. Then, 

iia) If  $S^A (i,j) = S^B (i,j)$, which means that the selected site is not
damaged, the Glauber dynamics is followed according to the standard 
procedure \cite{herr}. However,

iib) If $S^A (i,j) \neq S^B (i,j)$, i.e. for a damaged site, a new
random number $h$ is generated and one proceeds as follows:

iib1) If $h < q$, the damage is healed, so that one sets
$S^A (i,j)=S^B (i,j)$.  However, 

iib2. If $h \geq q$, the standard dynamics is applied.

According to the defined rules and taking $p = 1 -q$, one has
that for $p \rightarrow 0$, the damage would become healed for 
all temperatures, while in the limit $p \rightarrow 1$, the usual damage spreading dynamics in the Ising model is recovered.

Using the above described procedure, we have followed the time evolution 
of the damage $D(t)$, which is evaluated according to equation (\ref{eq:dam}),
for different values of $p$, $1.1 \leq T \leq 50$ and the lattice 
size $50 \leq L \leq 1000$. Simulations are stopped when the damage 
is healed, otherwise they are performed up to $t = 65000$ mcs.

\section*{III. Results and discussion}

Figure 1 shows log-log plots of $D(t)$ versus $t$ obtained at $T = 2$
for different values of $p$. It is found that: i) for \( p > p_{c} \) 
the damage becomes healed for finite times and the log-log plots of $D(t)$
versus $t$ exhibit a downward curvature. ii) For \( p <p_{c} \) 
the damage quickly propagates and the log-log plots of $D(t)$ versus $t$ 
show that the damage reaches a stationary value. 
Finally, just at \( p = p_{c} = 0.1895(5) \)
the damage decays according to a power law. So, the change of the
curvature of the plots shown in figure 1 allows us to identify the critical
probability \( p_{c} \) for damage spreading.

The straight line observed in figure 1 for \( p_{c} \) suggests 
a power-law behavior given by

\begin{equation}
D(t)\propto t^{-\delta}
\label{pwlaw}
\end{equation}

\noindent where \( \delta \) is an exponent. The best fit of the data 
gives \( \delta = 0.450 \pm 0.005 \). Similar plots obtained for $T > T_{C}$
(not shown here for the sake of space) exhibit the same behavior as that
observed in figure 1 and the typical slope, averaged within the range
$2 \leq T \leq 50$, is given by \( \delta = 0.45 \pm 0.01 \). 

On the other hand, for $p > p_{c}$ figure 1 shows a fast decay
of the damage. As suggested by the study of spreading 
behavior \cite{gradela}, we proposed an exponential decay given by

\begin{equation}
D(t) \propto exp(-t/ \tau)
\label{expodeca}
\end{equation}

\noindent where $\tau$ is the characteristic time of decay.

In order to check the scaling Ansatz given by equation (\ref{expodeca}), 
figure 2 shows log-lineal plots of $D(t)$ versus $t$ obtained for
$p > p_{c}$. The obtained straight lines for the long time 
behavior allow us to evaluate the dependence of $\tau$ on $p$.

It is found that $\tau$ increases when approaching the critical probability.
Furthermore, plots performed at different temperatures (no shown here for the 
sake of space) also exhibit the same behavior, independent of $T$.
Again, using the spreading formalism \cite{gradela}, the following 
power-law can be proposed

\begin{equation}
\tau \propto (p-p_{c})^{-\nu_{\parallel}}
\label{nu}
\end{equation}

\noindent where $\nu_{\parallel}$ is the correlation length exponent
for damage propagation along the time direction. Figure 3 shows log-log
plots of $\tau$ versus $p - p_{c}$ obtained at different 
temperatures. The typical averaged value resulting from the fits
is given by $\nu_{\parallel} = 1.25(5)$.

It is worth mentioning that our estimations for the exponents, given by 
$\delta = 0.45(1)$ and $\nu_{\parallel} = 1.25(5)$ are in agreement with
the well known critical exponents of the universality class of 
Directed Percolation (DP), namely $\delta^{DP} = 0.4505(10)$ \cite{h34}
and $\nu_{\parallel}^{DP}=1.295(6)$ \cite{f33}. So, these results strongly 
suggest that the the proposed model of damage spreading with damage healing
belongs to the DP universality class. This statement is further supported
by the fact that the limit $p \rightarrow 1$ corresponds to the well known
Ising model where the damage spreading transition is known to lie 
within the DP universality class \cite{grass2,grass3}.

On the other hand, as follows from figure 1, 
for $p < p_{c}^{D}$ and after a short decay period
($t \approx 10^{2}$ mcs), the damage becomes stabilized (or saturated) 
with typical average values $D_{sat}$ that depend on $p$. This trend
has been checked by means of extensive simulations run  
up to $t=10^6$ mcs. Furthermore, we have also checked that
the stationary values of $D_{sat}$ exhibit negligible finite-size effects for 
$L \ge 200$.  Figure \ref{Fig4} shows the dependence of $D_{sat}$ 
on $(p_{c}^{D}-p)$ in a lineal-lineal plot. It follows that 
$D_{sat}$ increases lineally as a function of $(p_{c}^{D}-p)$, 
and that the slope $F(T)$ depends on temperature.
%#### AGREGADO
This assumption is valid only in the limit of $p\rightarrow p_{c}^{D}$.
%###
Accordingly, we propose

\begin{equation}
D_{sat} = D_0 + F(T) (p_{c}^{D} - p) ,
\label{dsat}
\end{equation}

\noindent where $D_0 \approx 0.05$ is almost a constant independent of
$T$ (see figure \ref{Fig4}). The slope of the lines can be
obtained by fitting the data of figure \ref{Fig4} and it is found that
$F(T)$ also depends lineally with $T$  (see figure \ref{Fig5}) according to 

\begin{equation}
F(T)=F_0 + A \left( \frac{T-T_C}{T_C}\right)
\label{fdeT}
\end{equation}

\noindent where  $F_0 \approx 0$ and  $A \approx 1.5$ are also constants.

So, by inserting equation (\ref{fdeT}) into equation (\ref{dsat}),
one obtains

\begin{eqnarray}
D_{sat} = D_0 + \left( F_0 + A \frac{T-T_C}{T_C)}\right) \cdot (p-p_{c}^{D}) ,
\label{Dsatscal}
\end{eqnarray}

\noindent that gives the full dependence of $D_{sat}$ on both $p$ and $T$.
The scaling plot suggested by equation (\ref{Dsatscal}) holds acceptably as
shown in figure \ref{Fig6}. 

\begin{figure}
\centerline{{\epsfysize=5in \epsffile{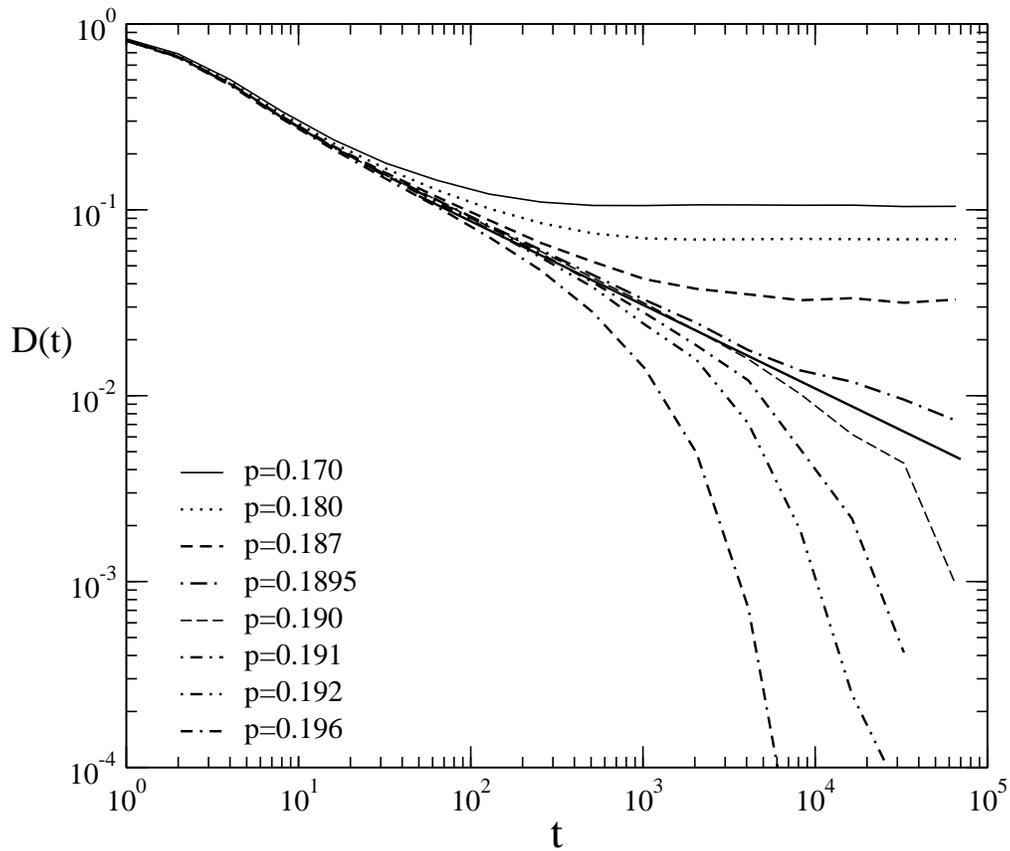}}}
\vskip 1.0 true cm
\caption{Log-log plots of the damage ($D(t)$) versus time $t$. Data obtained
using lattices of side $L=400$, for $T = 2.0T_C$ and taking different values
of $p$, as listed in the figure. The full line has slope 
$-\delta = -0.45$ (see equation (\ref{pwlaw}) ).}
\label{Fig1}
\end{figure}

Performing plots of the time evolution of the damage using different values of 
$p$ (as shown in figure \ref{Fig1}) and changing $T$, it is finally possible 
to evaluate the phase diagram of the model, i.e. a plot of $p_{c}^{D}$ as a 
function of $(T-T_C)/T_C$, as shown in figure \ref{Fig7}. It is found that
$p_{c}^{D} \rightarrow 0$  for very large values of $T$, as expected from the 
definition of the model. For the asymptotic decay of $p_{c}^{D}$ we propose

\begin{equation}
p_{c}^{D} \propto \left( \frac{T-T_C}{T_C}\right) ^{-\alpha}
\label{fasedia}
\end{equation}

\noindent where the best fit of the data, within the range 
$T \gg 1.5 T_C \sim 3.4J$, gives $\alpha=1$ for the exponent of equation
(\ref{fasedia}). It is worth mentioning that, for the lattices used in the
simulations, the phase diagram exhibits negligible deviations due to
finite size effects, as shown in figure \ref{Fig7}. On the other hand, 
close to $T_C$, it is observed that $p_{c}^{D}$ reaches a maximum value of 
the order of $p_{c}^{D} \sim 0.22$ for  $ T = 1.5 T_C \sim 3.4J$. We have 
carefully checked that this behavior, which remains to be understood,
is not due to the operation of finite size effects.
Finally, let us remark that all phase transitions at 
the critical curve at the border between
damage healing and damage spreading, shown in figure \ref{Fig7},
belong to the universality class of directed percolation.
 
\section*{IV. Conclusions}

Based on the fact that in the 2D Ising model the damage 
is not healed above the critical temperature, we proposed 
a model that introduces a healing probability. It is found 
that for $T > T_{C}$ the damage becomes healed for critical 
values of $p$. The phase diagram of the model is obtained and 
it is shown that the transitions between states of damage 
spreading and healing belong to the universality class of 
directed percolation in $(2 + 1)-$dimensions.
Our results thus support the conjecture that damage spreading
transitions are generically in the universality class of directed
percolation \cite{grass2}, provided the fact that such transitions 
do not coincide with the critical point of the undamaged system.

\newpage

\newpage
\begin{figure}
\centerline{{\epsfysize=5in \epsffile{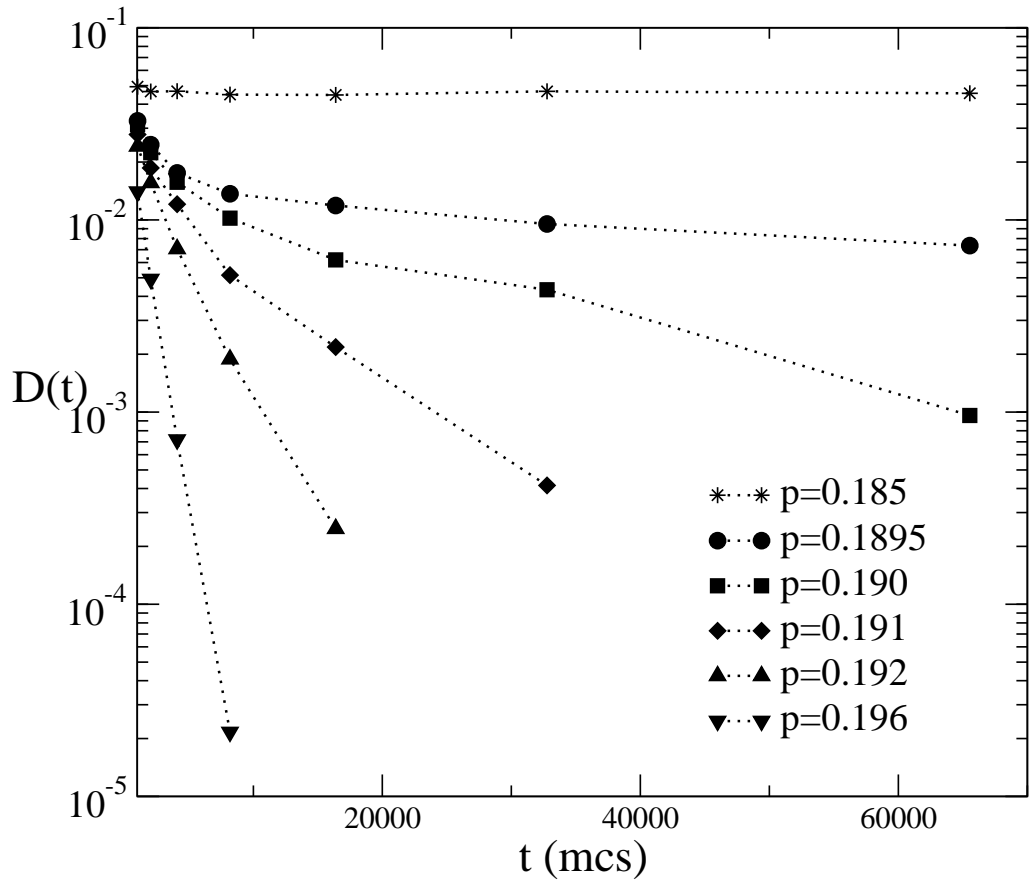}}}
\vskip 1.0 true cm
\caption{Log-lineal plots of the damage ($D(t)$) versus $t$ for data 
corresponding to $p > p_{c}^{D}$ and obtained taken $L=400$ and 
$T=2.0T_C$.}
\label{Fig2}
\end{figure}

\newpage

\begin{figure}
\centerline{{\epsfysize=5in \epsffile{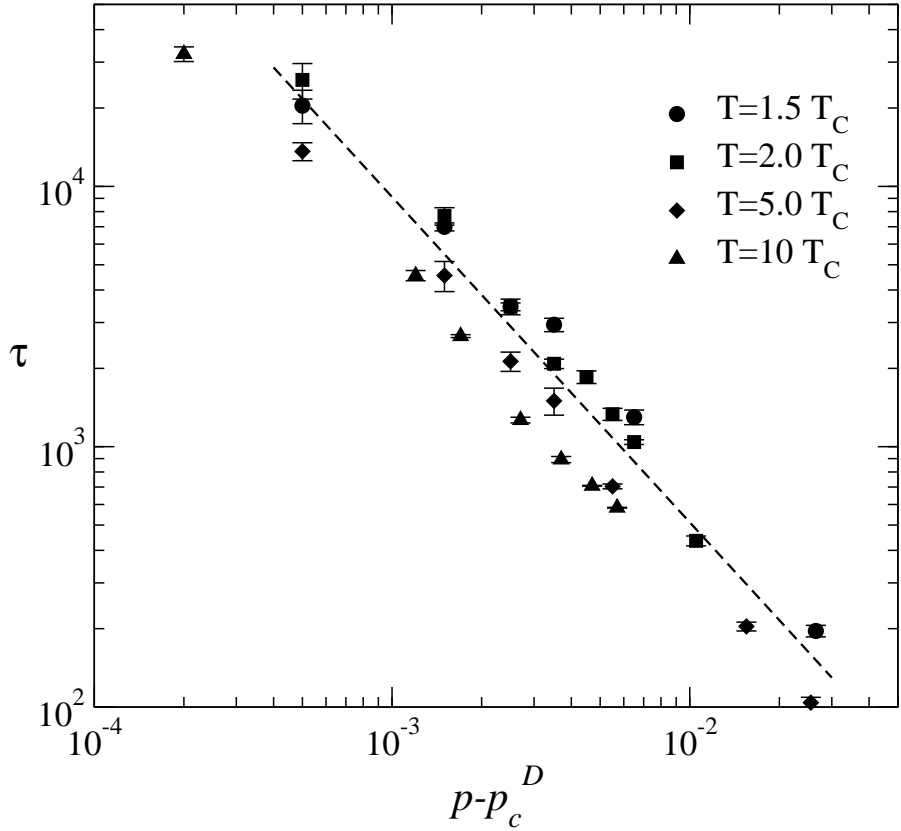}}}
\vskip 1.0 true cm
\caption{Log-log plot of characteristic time for damage healing $\tau$, 
obtained by fitting equation (\ref{expodeca}) to data as shown in figure 
\ref{Fig2}, versus $p-p_{c}^{D}$. Data obtained using lattices of size 
$L = 400$ and different temperatures, as listed in the figure. 
The dashed line, which has slope $\nu_{\parallel} = 1.25 $, is the
slope obtained averaging over all measured temperatures and has been 
drawn for the sake of comparison.}
\label{Fig3}
\end{figure}

\newpage

\begin{figure}
\centerline{{\epsfysize=5in \epsffile{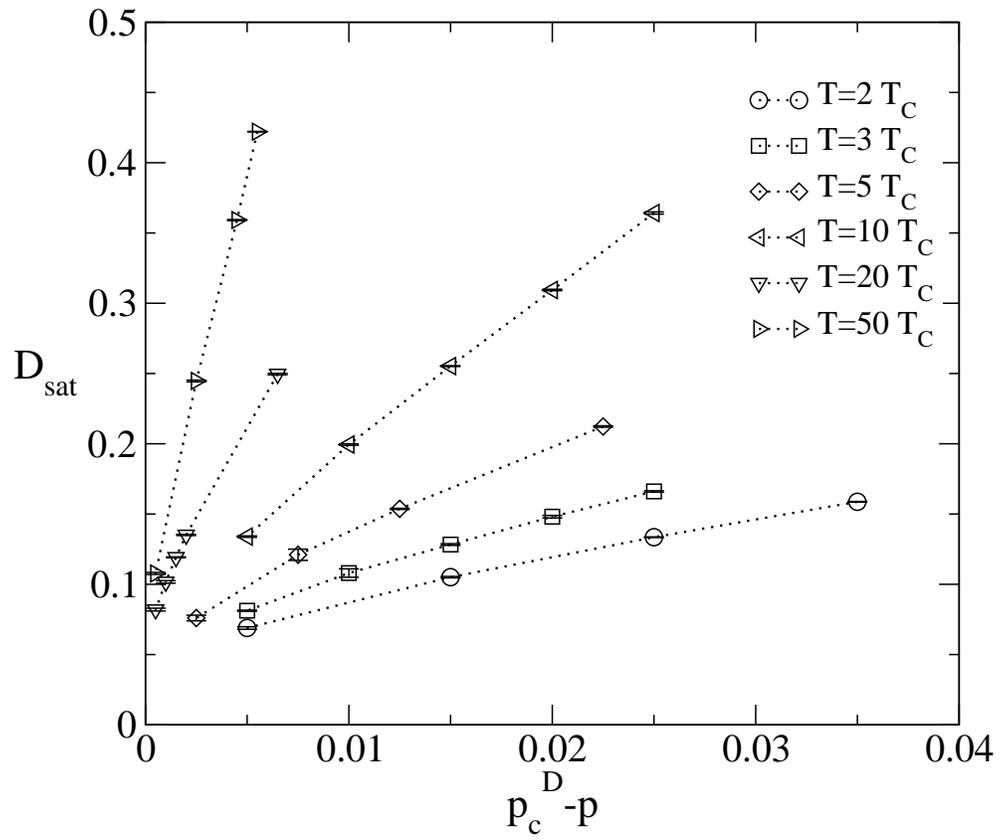}}}
\vskip 1.0 true cm
\caption{Lineal-lineal plots of $D_{sat}$ versus $p_{c}^{D}-p$ obtained for
different values of $T$, as listed in the figure.}
\label{Fig4}
\end{figure}

\newpage 

\begin{figure}
\centerline{{\epsfysize=5in \epsffile{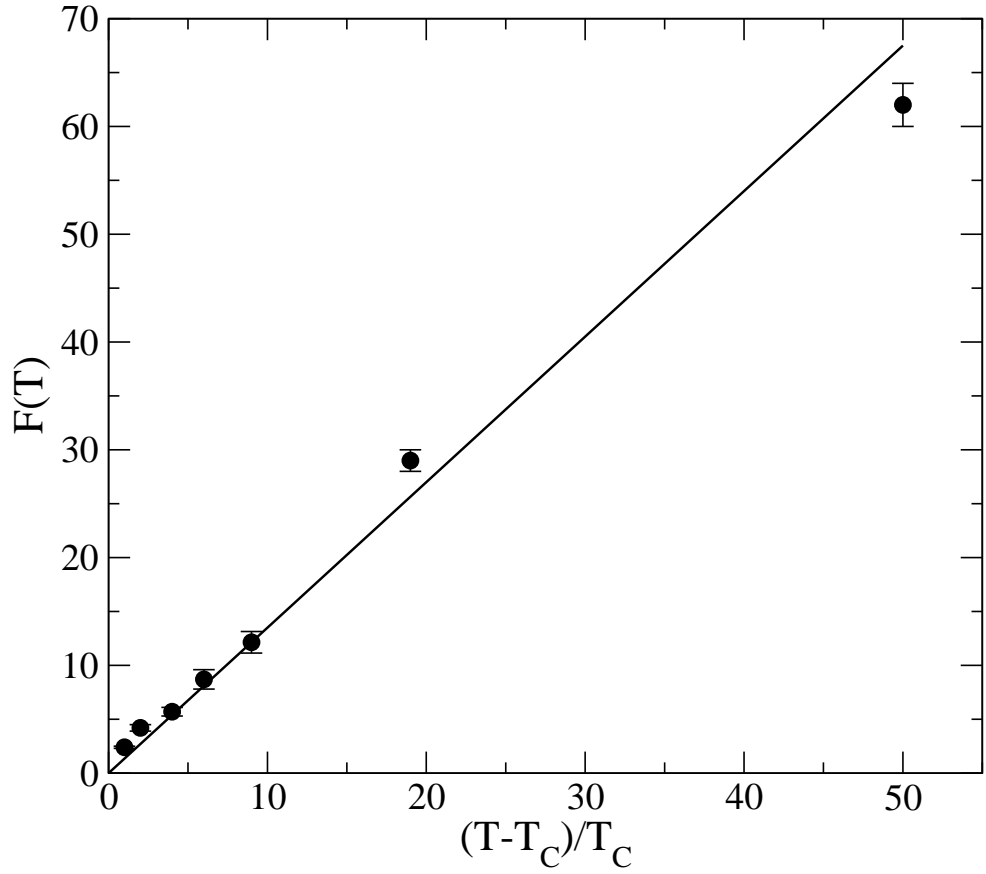}}}	
\vskip 1.0 true cm
\caption{Lineal-lineal plots of the slopes obtained by fitting the data shown
in figure \ref{Fig4} by means of equation (\ref{dsat}) given by $F(T)$, 
versus $(T-T_C)/T_C$. The straight line with slope $A \approx 1.5$ was 
obtained by fitting the data according to equation (\ref{fdeT}).}
\label{Fig5}
\end{figure}

\newpage

\begin{figure}
\centerline{{\epsfysize=5in \epsffile{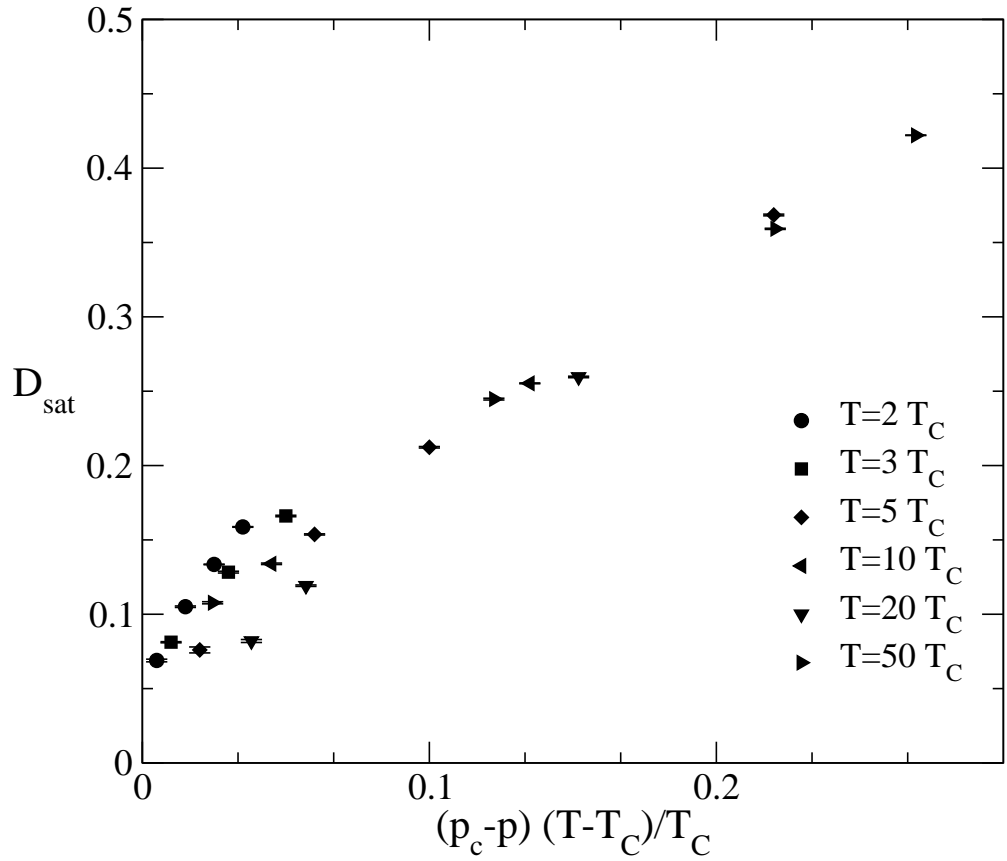}}}
\vskip 1.0 true cm
\caption{Scaling plot of $D_{sat}$ according to equation (\ref{Dsatscal}),
obtained taking $F_{0} = 0$. More details in the text.}
\label{Fig6}
\end{figure}

\newpage

\begin{figure}
\centerline{{\epsfysize=7in \epsffile{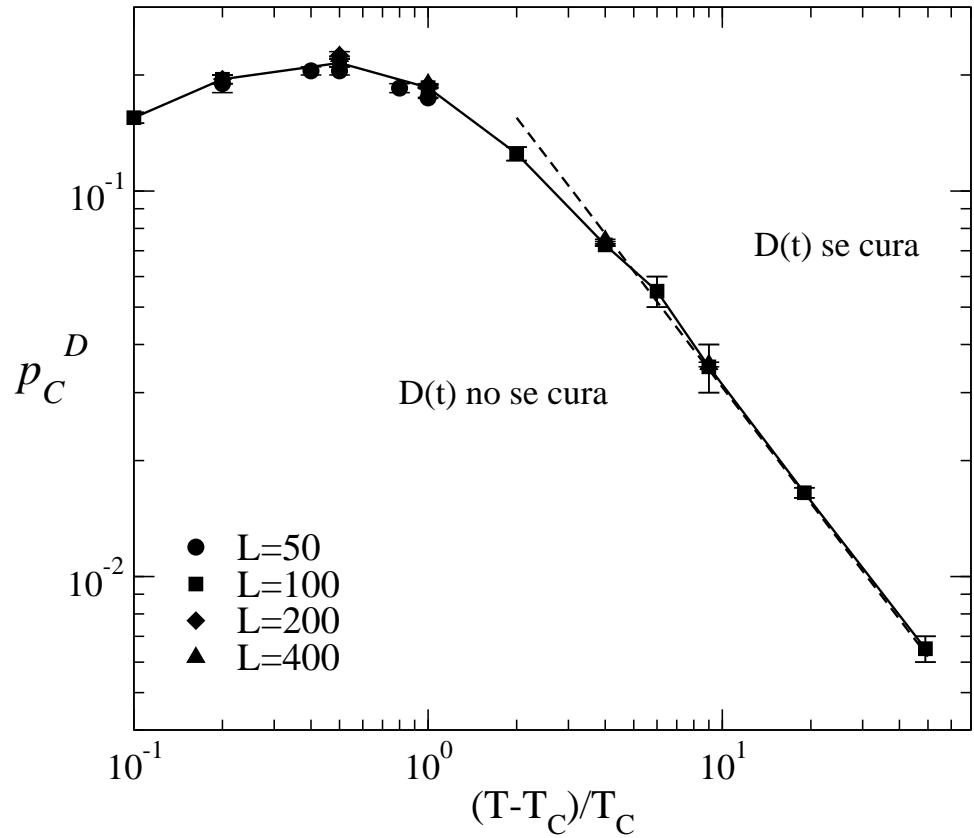}}}
\vskip 1.0 true cm
\caption{Phase diagram for the proposed model for damage spreading with 
a probability of damage healing. The full line has been drawn to guide 
the eyes.
The dashed line corresponds to the fit of the data for $T \gg 1.5 T_C$
and has slope $-\alpha = -1$.} 
\label{Fig7}
\end{figure}

\end{document}